\documentclass[epjCONF,onecolumn]{svjour}
\usepackage{graphics}
\usepackage[varg]{txfonts} 
\usepackage[latin1]{inputenc}
\session-title{Hot and Cold Baryonic Matter -- HCBM 2010}
\begin{document}
\title{Deflected Jets or Hot Spots? Conical Correlations of Hard
Trigger Particles}
\author{Barbara Betz\inst{1}\fnmsep\thanks{\email{betz@phys.columbia.edu}}}
\institute{ $^1$ Department of Physics, Columbia University, 
New York, 10027, USA}
\abstract{
The double-peak structure observed in soft-hard dihadron correlations 
was recently studied intensively in order to learn more about the 
jet-induced medium excitation in ultrarelativistic heavy-ion collisions.
Experimental data shows that the double-peak structure obtained for 
soft trigger particles coalesces into one peak for harder trigger particles.
We demonstrate that this effect occurs when averaging over many jet events in a 
transversally expanding background, while a hot spot scenario always leads to two distinct
peaks. This suggests to study soft-hard correlations induced by 
heavy-flavor jets with those generated by light-flavor jets at RHIC 
and LHC in order to really disentangle medium effects from jets.
} 
\maketitle
%


The hot and dense medium created in ultrarelativistic heavy-ion 
collisions \cite{whitebrahms,whitephenix,whitephobos,whitestar,Aamodt:2010pb,Aamodt:2010pa},
which is most likely the quark-gluon plasma (QGP), can be probed
with the help of jets. It is assumed that those jets are created
in the early stages of the collisions and interact with the
expanding system.
At RHIC it was found that the medium behaves as a nearly perfect 
fluid \cite{Romatschke:2007mq} and that it is opaque to jets 
\cite{Gyulassy:2004zy}, like at the LHC \cite{Aamodt:2010jd}. This
raises the possibility of studying medium properties using the
correlations of soft and hard particles. 

The interest in the experimental multi-particle correlations 
\cite{Adams:2005ph,Adler:2005ee,2pcPHENIX,UleryPRL} is based on the
double-peak structure found at angles opposite to the trigger jet,
which has been suggested as a signal for the creation of Mach cones
\cite{Stoecker:2004qu,CasalderreySolana:2004qm}.

In a fluid with low viscosity, Mach cones are generated by the interference
of sound waves resulting from the energy deposited by a supersonic
jet. They should lead to an excess of low-$p_T$ hadrons which are
emitted at an angle $\pi - \phi_M$ with respect to the
trigger jet. The Mach-cone angle $\phi_M$ is given by Mach's
law, $\cos\phi_M= c_s/v_{\rm jet}$, providing a possibility to
extract the speed of sound $c_s$.

Experimental multi-particle correlations were studied
intensively. It was shown that the position of the away-side
peaks does not change with $p_T^{assoc}$ (excluding Cherenkov gluon
radiation as a source for the double-peak structure), but strongly 
depends on $p_T^{trig}$. While a clear double-peak structure is seen 
for smaller $p_T^{trig}$ ($3<p_T^{trig}<4$~GeV), this structure coalesces
into one peak for larger $p_T^{trig}$ ($6<p_T^{trig}<10$~GeV) 
\cite{:2008cqb,Aggarwal:2010rf}. 

Recently, however, it has been shown \cite{Takahashi:2009na} 
that the experimentally observed two-peak structure for small 
$p_T^{trig}$ and $p_T^{assoc}$ can also be
obtained in two-particle correlations without considering jets, but
hot spots which occur due to the fluctuation of initial conditions. 
The flow created by the hot spot will interfere with the flow of
the expanding medium, forming a conical structure and resulting
in two peaks on the away-side.

Moreover, it was suggested in Ref.\ \cite{Alver:2010gr} that the
triangular flow ($v_3$) could also lead to the away-side features
observed in experimental data. This effect was studied in detail, both
experimentally \cite{Agakishiev:2010ur} and theoretically 
\cite{Xu:2010du,Ma:2010dv}. Unfortunately, it seems to yet remain inconclusive
if the conical structure will still be present after the subtraction 
of the triangular flow. This question can only be resolved after 
an experimentally extracted $v_3$ component is subtratced from the 
measured data.

In the following, however, we will demonstrate an effect that might lead, 
for very central events, to a weakening of the double-peak structure 
at larger $p_T^{trig}$, considering jets traversing through the medium. 
It will clearly differ from a hot spot
event (as presented below) and is closely connected to the path 
lengh dependence of a jet \cite{Sickles:2008uh,Afanasiev:2007tv}. 

Previous calculations 
\cite{CasalderreySolana:2006sq,Chaudhuri:2005vc,Renk:2006mv,Neufeld,Gubser:2007ga,Noronha:2008un,Betz:2008wy,Molnar:2009kx,Bouras:2010jd,Betz:2008ka} 
have shown that the formation of a conical structure on the away-side 
of soft-hard correlations can be very sensitive to the underlying 
assumptions about the jet-medium interaction \cite{Torrieri:2009mv}. 
While in case of a static medium a diffusion wake moving in the opposite 
trigger-jet direction may overwhelm any signal from the Mach cone leading 
to a single peak on the away-side \cite{CasalderreySolana:2006sq,Betz:2008wy,Molnar:2009kx,Bouras:2010jd,Betz:2008ka}, 
the strong longitudinal and transverse expansion of the QGP can distort 
the Mach-cone signal \cite{Renk:2006mv,Satarov:2005mv,Betz:2010qh}. 
This diffusion wake is universal to strongly and weakly-coupled energy 
loss \cite{Betz:2008wy}.

Assuming that the energy lost by the jet thermalizes quickly \cite{Adams:2005ph},
we solve the conservation equations
\begin{eqnarray}
\partial_\mu T^{\mu\nu}&=&S^\nu\,,
\end{eqnarray}
of the energy-momentum tensor 
$T^{\mu \nu} = (e+p) u^\mu u^\nu - p\, g^{\mu \nu}$, where $u^\mu$ is
the four-velocity of the fluid, using the $(3+1)$-dimensional hydrodynamic 
SHASTA algorithm \cite{Rischke:1995pe} for an ideal gas EoS ($p=e/3$) of 
massless $SU(3)$ gluons. $S^\nu$ denotes the energy and momentum
deposited by a jet. We choose the following ansatz
\begin{eqnarray}
\label{SourceExpandingMedium0}
S^\nu (x) = \int\limits_{\tau_i}^{\tau_f}d\tau 
\frac{dM^\nu}{d\tau} \, \frac{u_\alpha j^\alpha}{u_{0,\beta} j^\beta_0}\,
\delta^{(4)} \left[ x - x_{\rm jet}(\tau) \right],
\end{eqnarray}
with the proper-time interval of the jet evolution $\tau_f - \tau_i$,
the (constant) energy and momentum loss rate $dM^\nu/d\tau=(dE/d\tau,
d\vec{M}/d\tau)$, and the location of the jet $x_{\rm jet}$.
Here, $j^\alpha$ is the four-current of color charges and
$u_0^\beta$, $j_0^\beta$ are the initial four-velocity and
four-current of color charges at the center of
the system, respectively. Thus, the factor 
$u_\alpha j^\alpha/(u_{0,\beta} j^\beta_0)$
takes into account that the medium expands and cools, 
reducing the energy-momentum loss rate. In non-covariant notation, 
Eq.\ (\ref{SourceExpandingMedium0}) reads
\begin{eqnarray}
\label{SourceExpandingMedium}
S^\nu(t,\vec{x}) &=& \frac{1}{(\sqrt{2\pi}\,\sigma)^3}
\exp\left\{ -\frac{[\vec{x}-\vec{x}_{\rm jet}(t)]^2}{
2\sigma^2}\right\} 
 \left(\frac{dE}{dt},\frac{dM}{dt},0,0\right)
\left[\frac{T(t,\vec{x})}{T_{\rm max}}\right]^3\,.
\end{eqnarray} 
In the following, $\sigma=0.3$~fm.

We investigate an expanding medium with an initial transverse
energy density profile given by the Glauber model 
for a maximum temperature of either $T=200$~MeV (Au+Au) or $T=176$ MeV 
(Cu+Cu). Note that the exact value of the initial temperature
does not play an important role for the analysis since we are 
considering an ideal gas EoS. In the longitudinal direction, the system is
assumed to be a cylinder, elongated over the whole grid. With this assumption, we
minimize the effect of longitudinal flow.  A temperature cut of
$T_{\rm cut}=130$~MeV is applied 
to ensure that no energy-momentum deposition takes place outside the 
medium. Since it was shown in Ref.\ \cite{Betz:2008ka} that jet
deceleration does not lead to significant changes in the particle
correlations after freezeout, we consider that the jets move at
a constant velocity through the expanding medium. 

However, we assume that each parton moving through the QGP will
eventually be completely thermalized after the deposition of all its
initial energy. This is an important difference to the ansatz chosen
in Ref. \cite{Chaudhuri:2005vc} where the jet was energetic enough
to punch through the medium.

Here we consider the jet to be generated by a $5$~GeV or 
$11.4$~GeV parton which corresponds to a trigger-$p_T$ of 
$3.5$~GeV and $8$~GeV, respectively, assuming that, after 
fragmentation, the leading hadron carries $\sim 70$\% of the parton's 
energy.

Since experiment can trigger on the jet direction, but not on the location
where the jet was formed, one has to consider different jet trajectories 
pointing along the same direction but originating from different points
in the transverse plane \cite{Chaudhuri:2006qk}. 
We parametrize these trajectories as
\begin{eqnarray}
\label{paths}
x &=& r\cos\phi \hspace*{1.5cm} y = r\sin\phi\,,
\end{eqnarray}
where $r=5$~fm is chosen to account for surface bias and consider
different values for the azimuthal jet angle with respect to the trigger
axis (which is chosen to be the negative $x$-axis). Here we
denote all jets travelling at $\phi=90, \ldots, 165$~degrees (with
a $\Delta\phi=15$~degrees) as jets propagating through the
upper half of the medium and jets between $\phi = 195, \ldots, 270$~
degrees as those going through the lower half of the medium.

After the hydrodynamic evolution, the fluid is converted into
particles using the Cooper-Frye (CF) prescription \cite{Cooper:1974mv} 
at a constant time (isochronous freeze-out) which leads to the 
single-inclusive particle spectrum $dN/(p_T dp_T dy d\phi)$.

\begin{figure*}[t]
\centering
\begin{minipage}[c]{4.2cm}
\resizebox{1.2\columnwidth}{!}{\includegraphics{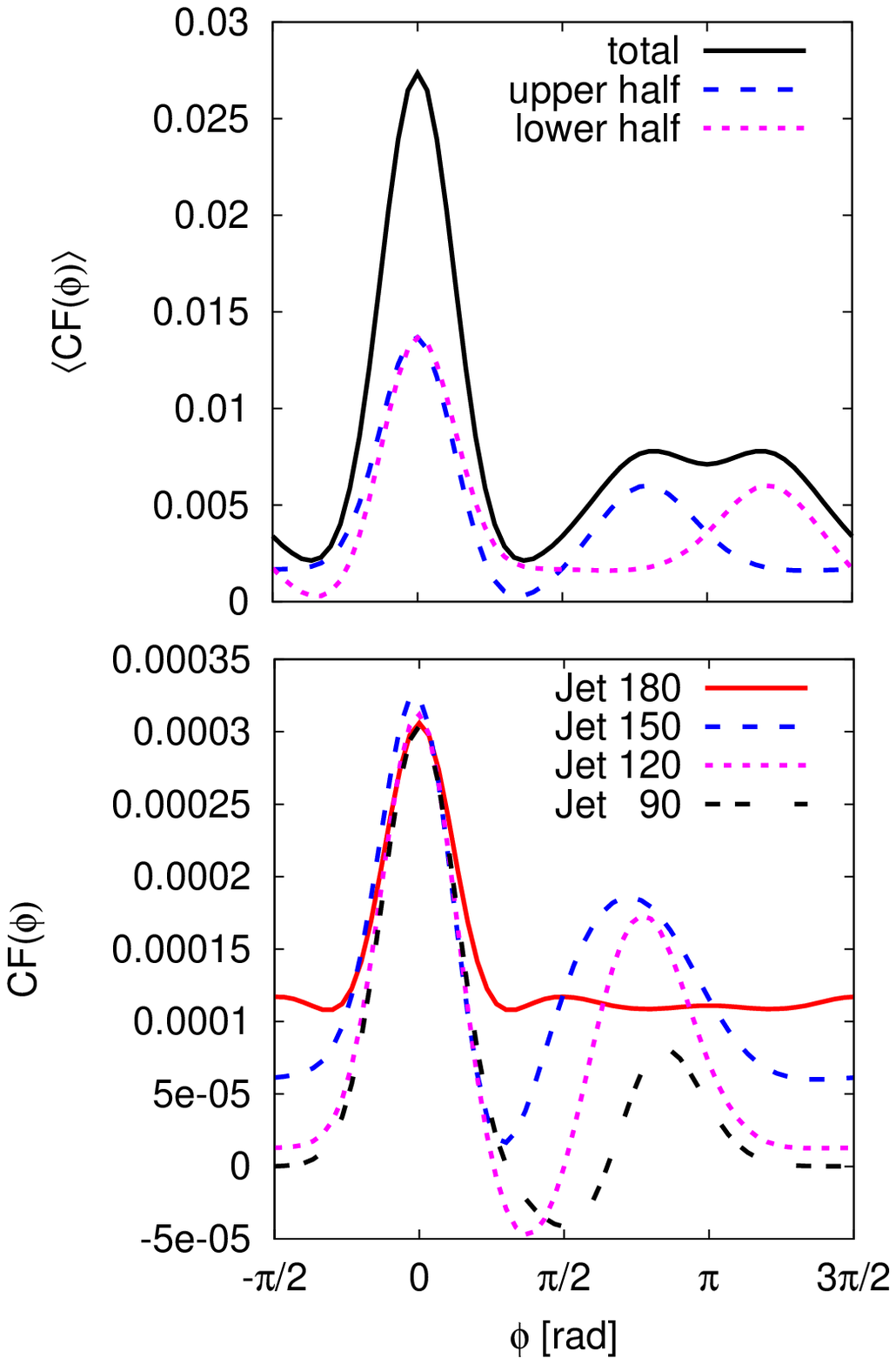}}
\end{minipage}
\hspace*{0.9cm}  
\begin{minipage}[c]{4.2cm}   
\resizebox{1.2\columnwidth}{!}{\includegraphics{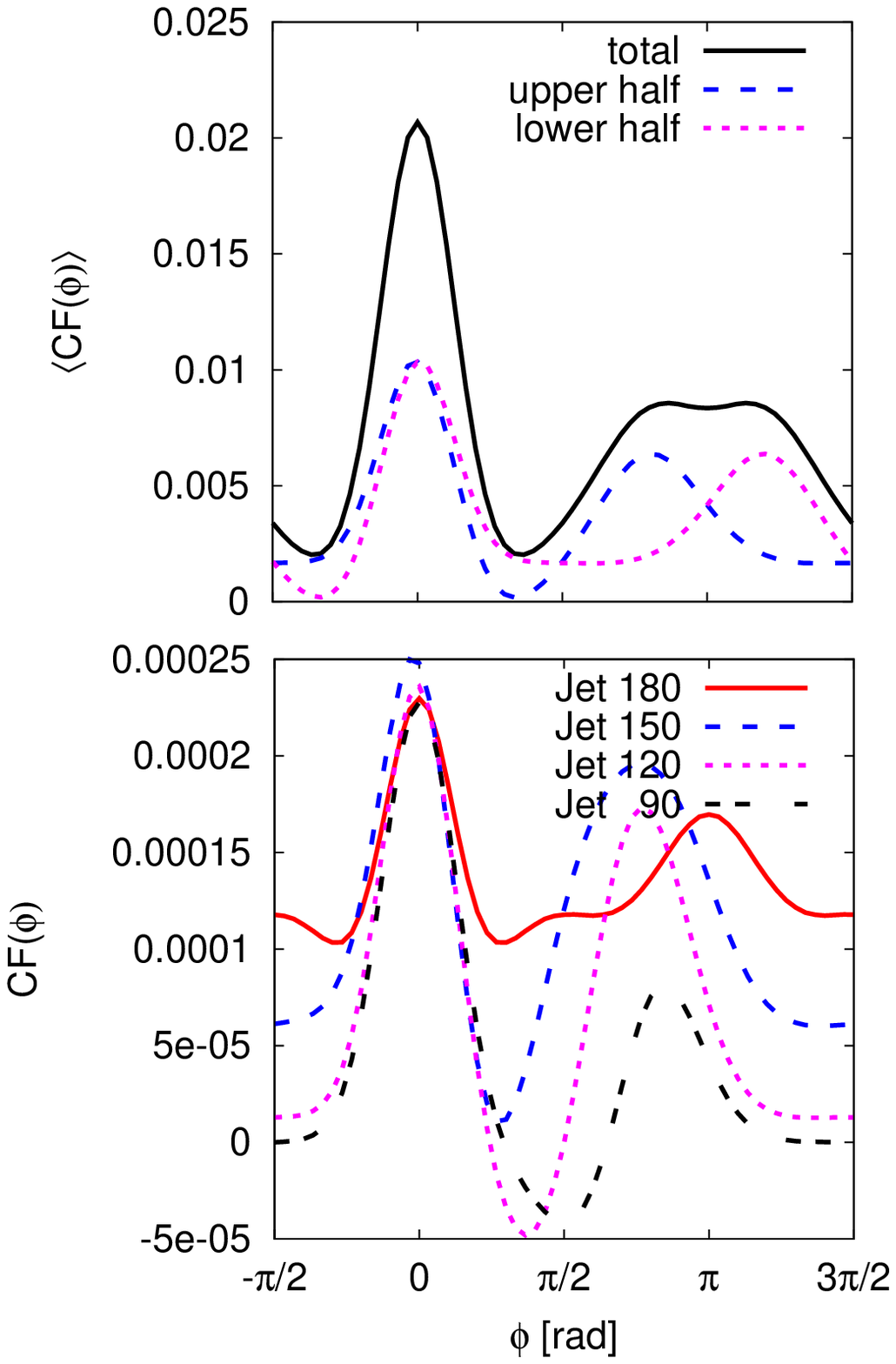}}
\end{minipage}
  \caption{The two-particle correlation function 
   (solid black line) for a $p_T^{trig}=3.5$~GeV (left panel) 
   \cite{Betz:2010qh} and for a $p_T^{trig}=8$~GeV (right panel), assuming
   that the associated particle is $p_T^{assoc}=2$~GeV.
   The long-dashed blue and short-dashed magenta lines in the upper panels 
   represent the averaged contribution from jets traversing only 
   the upper or the lower half of the medium, respectively. The
   unaveraged two-particle correlation function is shown
   in the lower panels from four representatively chosen different jet 
   trajectories in the upper half of the medium.}
  \label{Fig1a}
\end{figure*}

One major difference between the experimental situation and the 
hydrodynamical calculation proposed above is that the trajectory of the jet
is not known in the first case. Thus, one has to measure
the azimuthal correlation between hard particles produced by the
trigger jet and soft particles produced by the associated jet. 
We mimic the hard-soft correlation function by convoluting the
single-inclusive particle spectrum (which only considers the 
away-side particles) with a function representing the near-side jet,
\begin{eqnarray}
f(\phi) &=& \frac{1}{\sqrt{2\pi\Delta\phi^2}}
\exp\left(-\frac{\phi^2}{2\Delta\phi^2}\right)\,,
\end{eqnarray}
($\Delta\phi=0.4$), resulting in a 
two-particle correlation function
\begin{eqnarray}
C_2(\phi)&=& A f(\phi) + \int\limits_0^{2\pi}
d\phi^\star \frac{dN}{p_T dp_T dy d(\phi-\phi^\star)} f(\phi^\star)\,,
\end{eqnarray}
where $(A,\Delta\phi)$ are chosen to simulate the  
near-side correlation.  This function is then event-averaged (indicated by
$\langle \cdot \rangle$), background-subtracted, and 
normalized, leading to the averaged two-particle correlation function
\begin{equation} \label{CF}
\langle CF(\phi)\rangle= {\cal N}\,
\left[\langle C_2(\phi) \rangle - 
\frac{d N_{\rm back}}{p_T dp_T dy d\phi}
\right]\,,
\end{equation}
where $d N_{\rm back}/(p_T dp_T dy d\phi)$ is the single-inclusive
particle spectrum for an event without jets and
${\cal N}^{-1} = d N_{\rm back}/(p_T dp_T dy)$.

\begin{figure}[t]
\centering
\resizebox{0.45\columnwidth}{!}{\includegraphics{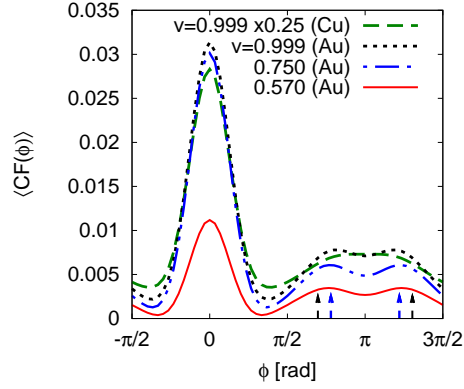}}  
  \caption{The two-particle correlation function, Eq.\ (\ref{CF}), 
   for light jets travelling at $v=0.999$ in central Cu+Cu collisions 
   (long-dashed green line) as well as central Au+Au collisions 
   (short-dashed black line), and $b$-jet propagating at $v=0.75$ 
   (dash-dotted blue line) and $v=0.57$ (solid red line) for a 
   $p_T^{assoc}=2$~GeV. For the supersonic jets, the 
   arrows indicate the emission angles obtained by Mach's law. In
   case of the Cu+Cu, the double-peak structure only appears for
   larger $p_T^{assoc}$ due to thermal smearing \cite{Betz:2010qh}.}
  \label{FigAllVelocities}
\end{figure}

Figure \ref{Fig1a} shows the two-particle correlation function 
for a $p_T^{trig}=3.5$~GeV (left) and a $p_T^{trig}=8.0$~GeV (right) 
assuming a $p_T^{assoc} = 2$~GeV.
The jets are considered to propagate with $v=0.999$, depositing energy 
and momentum into the medium according to Eq.\ 
(\ref{SourceExpandingMedium}) with $dE/dt = 1$~GeV/fm and 
$dM/dt = 1/v\,dE/dt$. 
In both cases we observe a double-peak structure resembling a Mach-cone 
signal, but the peak-to-valley ratio is much larger for the larger 
$p_T^{trig}$. 
The cone-like signal is a consequence of the different contributions
of the jet trajectories that are shown in the lower panels of 
Fig.\ \ref{Fig1a}. Those jets traversing the upper half of the medium
add up to a peak at an angle smaller than $180$~degrees (long-dashed 
blue lines in the upper panels of Fig.\ \ref{Fig1a}), while the
contributions from the jets traversing the lower half of the medium
(short-dashed magenta lines in the upper panels of Fig.\ \ref{Fig1a})
lead to a peak at an angle larger than $180$~degrees. 
The gap between those two peaks depends on how much the transversally
expanding medium deflects the matter in the disturbances caused by the 
jet as well as on $p_T^{assoc}$. 

For a larger $p_T^{trig}$ the jet traversing the middle of the medium
(red line in the lower right panel of Fig.\ \ref{Fig1a}) may reach that part
of the medium where the background flow of the expanding system is
parallel to the flow created by the diffusion wake 
\cite{Betz:2010qh,Neufeld:2008eg}, enhancing its impact and causing 
a contribution opposite to the trigger jet which fills up the 
double-peak structure. This effect might be seen in the data 
\cite{:2008cqb,Aggarwal:2010rf}, leading to a two-peak structure
for small $p_T^{trig}$ and just one broad away-side peak for large 
$p_T^{trig}$.

Thus, the conical shape results from the averaging over many different
jet events in an expanding medium \cite{Betz:2010qh}. It even
appears for subsonic jets (see Fig.\ \ref{FigAllVelocities}) which
demonstrates that the effect cannot be due to a true Mach cone or
used to conclusively distinguish between different jet deposition 
scenarios \cite{Betz:2010qh}.

\begin{figure}[t]
\centering
\resizebox{0.9\columnwidth}{!}{\includegraphics{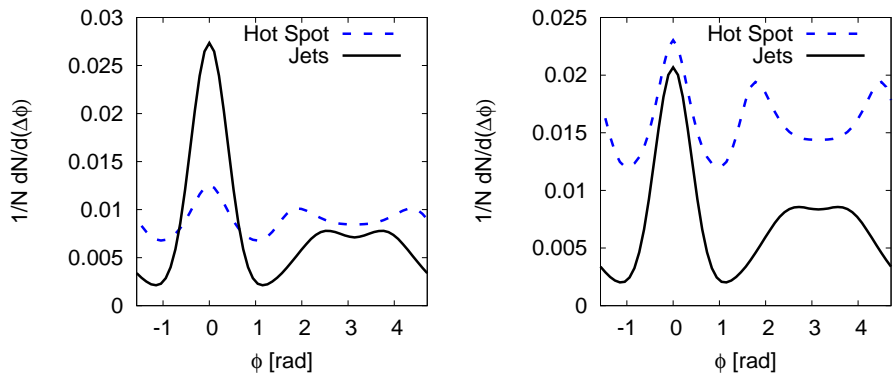}}  
  \caption{Two-particle correlation function from a hot spot event
  (dashed blue lines) and averaged jet events (black solid lines), 
  assuming a $p_T^{assoc}=2.0$~GeV and $p_T^{trig}=3.5$~GeV (left panel) 
  as well as $p_T^{trig}=8.0$~GeV (right panel).}
  \label{FigCorrelation}
\end{figure}

As discussed above, the two peak structure on the away side of
azimuthal correlations for small $p_T^{trig}$ and $p_T^{assoc}$
can also be obtained due to the evolution of a hot spot 
\cite{Takahashi:2009na}. However, the relevant question is if such a
double-peak structure also coalesces into one peak for larger $p_T^{trig}$.

To check this, we basically replaced a jet in the above setup of most central collisions
by a hot spot, choosing different $\Delta e/e_0$. Please note that the 
actual position of the hot spot is irrelevant in this case and it is
also not necessary to average over many events since there is no
trigger jet axis and thus all events can be
converted into each other due to rotational symmetry. 

Fig. \ref{FigCorrelation} shows the two-particle correlation function
for such a hot spot event (blue dashed lines) assuming a 
$p_T^{assoc}=2.0$~GeV and a $p_T^{trig}=3.5$~GeV (left panel) as well as
a $p_T^{trig}=8.0$~GeV (right panel), obtained according to
\begin{equation} \label{eqhotspot}
\frac{1}{N}\frac{dN}{d\Delta\phi}=
\frac{1}{N}\int\frac{dN}{d\phi}\frac{dN}{d(\Delta\phi-\phi)}d\phi
\end{equation}
and compared with the averaged jet events from Fig.\ \ref{Fig1a} 
(black solid line). Here we chose $\Delta e/e_0=6$, other ratios 
give similar results. 

As can be seen, the double-peak structure is more pronounced in 
case of a hot spot and gets even stronger for larger $p_T^{trig}$, 
in contrast to the averaged jet events where the two peak structure
starts to coalesce into one peak for larger $p_T^{trig}$ like seen in the data.

However, it seems rather unlikely that a hot spot creates particles
with very large $p_T$. But for small $p_T^{trig}$ 
\cite{:2008cqb,Aggarwal:2010rf} a superposition of 
fluctuating events (hot spots) and jets might probably lead to a rather
clean double-peak structure. This should also be seen at the LHC.

It is important to note in this context that the effect of triangular flow and hot
spots are very closely linked to each other and might actually not
be disentangled. Each hot spot (and thus fluctuating initial conditions) 
will lead to a nonzero triangular flow.


In conclusion, we have shown that a double-peak structure on the away side of
soft-hard correlations obtained via averaging over different jet events 
in which the particles are emitted from the deflected wakes created by jets
\cite{Betz:2010qh} coalesces into one peak for larger $p_T^{trig}$ as seen in 
experimental data \cite{:2008cqb,Aggarwal:2010rf}, in contrast to the
the two-peak structure obtained from hot spots. 
Since such a distinction is not possible experimentally, it is
necessary to study soft-hard correlations induced by heavy-flavor tagged 
jets \cite{Antinori:2005tu} with those induced by light-flavor jets at 
RHIC and LHC in order to disentangle the medium effects (hot spots) from 
jets and to test if a conical correlation occurs even for subsonic 
``jets'' \cite{Betz:2010qh}.
  
B.B.\ thanks J.\ Noronha, G.\ Torrieri, M.\ Gyulassy, and D.\ Rischke
for all the discussions and acknowledges support from the Alexander von 
Humboldt foundation via a Feodor Lynen fellowship and DOE under
Grant No.\ DE-FG02-93ER40764.


\begin{thebibliography}{62}

\vspace*{-0.3cm}

\bibitem{whitebrahms}
 I.~Arsene {\it et al.}  [BRAHMS Collaboration],
  Nucl.\ Phys.\ A {\bf 757}, 1 (2005).

\bibitem{whitephenix}
 K.~Adcox {\it et al.}  [PHENIX Collaboration],
  Nucl.\ Phys.\ A {\bf 757}, 184 (2005).

\bibitem{whitephobos}
  B.~B.~Back {\it et al.},
  Nucl.\ Phys.\ A {\bf 757}, 28 (2005).

\bibitem{whitestar}
  J.~Adams {\it et al.}  [STAR Collaboration],
  Nucl.\ Phys.\ A {\bf 757}, 102 (2005).

\bibitem{Aamodt:2010pb}
  K.~Aamodt {\it et al.}  [The ALICE Collaboration],
  arXiv:1011.3916 [nucl-ex].

\bibitem{Aamodt:2010pa}
  K.~Aamodt {\it et al.}  [The ALICE Collaboration],
  arXiv:1011.3914 [nucl-ex].

\bibitem{Gyulassy:2004zy}
  M.~Gyulassy and L.~McLerran,
  Nucl.\ Phys.\  A {\bf 750}, 30 (2005); E.~V.~Shuryak,
  Nucl.\ Phys.\  A {\bf 750}, 64 (2005).

\bibitem{Aamodt:2010jd}
  K.~Aamodt {\it et al.}  [ALICE Collaboration],
  arXiv:1012.1004 [nucl-ex].

\bibitem{Romatschke:2007mq}
  P.~Romatschke and U.~Romatschke, 
  Phys.\ Rev.\ Lett.\  {\bf 99}, 172301 (2007). 

\bibitem{Adams:2005ph}
  J.~Adams {\it et al.}  [STAR Collaboration],
  Phys.\ Rev.\ Lett.\  {\bf 95}, 152301 (2005).

\bibitem{Adler:2005ee}
  S.~S.~Adler {\it et al.}  [PHENIX Collaboration],
  Phys.\ Rev.\ Lett.\  {\bf 97}, 052301 (2006).

\bibitem{2pcPHENIX} 
  A.~Adare {\it et al.}  [PHENIX Collaboration], 
  Phys.\ Rev.\  C {\bf 78}, 014901 (2008).

\bibitem{UleryPRL}
  B.~I.~Abelev {\it et al.}  [STAR Collaboration], 
  Phys.\ Rev.\ Lett.\  {\bf 102}, 052302 (2009). 

\bibitem{Stoecker:2004qu}
  H.~Stoecker,
  Nucl.\ Phys.\  A {\bf 750}, 121 (2005).

\bibitem{CasalderreySolana:2004qm}
  J.~Casalderrey-Solana, E.~V.~Shuryak and D.~Teaney,
  J.\ Phys.\ Conf.\ Ser.\  {\bf 27}, 22 (2005)
  [Nucl.\ Phys.\  A {\bf 774}, 577 (2006)].

\bibitem{:2008cqb}
  A.~Adare {\it et al.}  [PHENIX Collaboration],
  Phys.\ Rev.\  C {\bf 78}, 014901 (2008).

\bibitem{Aggarwal:2010rf}
  M.~M.~Aggarwal {\it et al.}  [STAR Collaboration],
  Phys.\ Rev.\  C {\bf 82}, 024912 (2010).

\bibitem{Takahashi:2009na}
  J.~Takahashi {\it et al.},
  Phys.\ Rev.\ Lett.\  {\bf 103}, 242301 (2009).

\bibitem{Alver:2010gr}
  B.~Alver and G.~Roland,
  Phys.\ Rev.\  C {\bf 81}, 054905 (2010)
  [Erratum-ibid.\  C {\bf 82}, 039903 (2010)].

\bibitem{Agakishiev:2010ur}
  H.~Agakishiev {\it et al.},
  arXiv:1010.0690 [nucl-ex].

\bibitem{Xu:2010du}
  J.~Xu and C.~M.~Ko,
  arXiv:1011.3750 [nucl-th].

\bibitem{Ma:2010dv}
  G.~L.~Ma and X.~N.~Wang,
  arXiv:1011.5249 [nucl-th].

\bibitem{Sickles:2008uh}
  A.~Sickles,
  Eur.\ Phys.\ J.\  C {\bf 61}, 583 (2009).

\bibitem{Afanasiev:2007tv}
  S.~Afanasiev {\it et al.}  [PHENIX Collaboration],
  Phys.\ Rev.\ Lett.\  {\bf 99}, 052301 (2007).

\bibitem{CasalderreySolana:2006sq}
  J.~Casalderrey-Solana, E.~V.~Shuryak and D.~Teaney,
  arXiv:hep-ph/0602183.

\bibitem{Chaudhuri:2005vc}
  A.~K.~Chaudhuri and U.~Heinz,
  Phys.\ Rev.\ Lett.\  {\bf 97}, 062301 (2006).

\bibitem{Renk:2006mv}
  T.~Renk and J.~Ruppert, 
  Phys.\ Lett.\  B {\bf 646}, 19 (2007); 
  Phys.\ Rev.\  C {\bf 76}, 014908 (2007).

\bibitem{Neufeld} 
  R.~B.~Neufeld, B.~Muller and J.~Ruppert,
  Phys.\ Rev.\  C {\bf 78}, 041901 (2008).

\bibitem{Gubser:2007ga}
  S.~S.~Gubser, S.~S.~Pufu and A.~Yarom,
  Phys.\ Rev.\ Lett.\  {\bf 100}, 012301 (2008).

\bibitem{Noronha:2008un}
  J.~Noronha, M.~Gyulassy and G.~Torrieri, 
  Phys.\ Rev.\ Lett.\  {\bf 102}, 102301 (2009).

\bibitem{Betz:2008wy}
  B.~Betz, M.~Gyulassy, J.~Noronha and G.~Torrieri,
  Phys.\ Lett.\  B {\bf 675}, 340 (2009).

\bibitem{Molnar:2009kx}
  D.~Molnar,
  AIP Conf.\ Proc.\  {\bf 1182}, 791 (2009).

\bibitem{Bouras:2010jd}
  I.~Bouras {\it et al.},
  arXiv:1008.4072 [hep-ph].

\bibitem{Betz:2008ka}
  B.~Betz, J.~Noronha, G.~Torrieri, M.~Gyulassy, I.~Mishustin and D.~H.~Rischke,
  Phys.\ Rev.\  C {\bf 79}, 034902 (2009).

\bibitem{Torrieri:2009mv}
  G.~Torrieri, B.~Betz, J.~Noronha and M.~Gyulassy,
  Acta Phys.\ Polon.\  B {\bf 39}, 3281 (2008).

\bibitem{Satarov:2005mv}
  L.~M.~Satarov, H.~Stoecker and I.~N.~Mishustin, 
  Phys.\ Lett.\  B {\bf 627}, 64 (2005).

\bibitem{Betz:2010qh}
  B.~Betz, J.~Noronha, G.~Torrieri, M.~Gyulassy and D.~H.~Rischke,
  Phys.\ Rev.\ Lett.\  {\bf 105}, 222301 (2010).

\bibitem{Rischke:1995pe}
  D.~H.~Rischke, Y.~Pursun, J.~A.~Maruhn, H.~Stoecker and W.~Greiner,
  Heavy Ion Phys.\  {\bf 1}, 309 (1995).

\bibitem{Chaudhuri:2006qk}
  A.~K.~Chaudhuri, 
  Phys.\ Rev.\  C {\bf 75}, 057902 (2007).

\bibitem{Cooper:1974mv}
  F.~Cooper and G.~Frye,
  Phys.\ Rev.\  D {\bf 10}, 186 (1974).

\bibitem{Neufeld:2008eg}
  R.~B.~Neufeld, 
  Eur.\ Phys.\ J.\ C {\bf 62}, 271 (2009).

\bibitem{Antinori:2005tu}
  F.~Antinori and E.~V.~Shuryak,
  J.\ Phys.\ G {\bf 31}, L19 (2005).

\end{thebibliography}
\end{document}